\documentclass[12pt,superscriptaddress,floatfix,a4paper]{article}
\usepackage[T2A]{fontenc}
\usepackage[cp1251]{inputenc}
\usepackage[english]{babel}
\usepackage{amsmath,amssymb,amsfonts}
\usepackage{bm,latexsym,euscript,textcomp}
\usepackage{graphicx,color,graphics,caption,epsf,dcolumn}
\usepackage{natbib}
\usepackage{hyperref}
\newcommand{\beq}{\begin{equation}}
\newcommand{\eeq}{\end{equation}}
\newcommand{\pt}{\partial}

\begin{document}

\title{\Large \bf Examining the physical principles behind the motion of moist air: Which expressions are sound?}

\author{A. M. Makarieva$^{1,2}$, V. G.  Gorshkov$^{1,2}$, A. V. Nefiodov$^1$,\\ D.  Sheil$^{3,4,5}$,
A. D. Nobre$^6$, P. Bunyard$^7$, B.-L. Li$^2$}

\maketitle

$^1$Theoretical Physics Division, Petersburg Nuclear Physics Institute, 188300, Gatchina,
St. Petersburg, Russia;
$^2$XIEG-UCR International Center for Arid Land Ecology, University of California, Riverside 92521-0124, USA.
$^3$School of Environment, Science and Engineering, Southern Cross University, PO Box 157, Lismore, NSW 2480, Australia;
$^4$Institute of Tropical Forest Conservation, Mbarara University of Science and Technology, PO Box, 44, Kabale, Uganda;
$^5$Center for International Forestry Research, PO Box 0113 BOCBD, Bogor 16000, Indonesia;
$^5$Centro de Ci\^{e}ncia do Sistema Terrestre INPE, S\~{a}o Jos\'{e} dos Campos SP 12227-010, Brazil;
$^6$Lawellen Farm, Withiel, Bodmin, Cornwall, PL30 5NW, United Kingdom and University Sergio Arboleda, Bogota, Colombia;
$^7$XIEG-UCR International Center for Arid Land Ecology, University of California, Riverside 92521-0124, USA.

\begin{abstract}
The physical equations determining the motion of moist atmospheric air in the presence of condensation remain controversial. Two distinct formulations
have been proposed, published and cited. The equation of \citet[][J. Atmos. Sci. \textbf{59}: 1967--1982]{ba02} includes a term for a "reactive motion"
that arises when water vapor condenses and droplets begin to fall; according to this term the remaining gas moves upwards so as to conserve momentum.
In the equation of \citet[][J. Atmos. Sci. \textbf{58}: 2073--2102]{oo01} the reactive motion term is absent. Both equations contain a term for condensate
loading, but in the formulation of \citet{oo01} there are two additional terms. In some modern nonhydrostatic models of moist atmospheric circulation,
however, formulations have been mixed. Here we examine the contrasting equations for the motion of moist air. We discuss inconsistencies in the
application of Newton's second and third laws to an air and condensate mixture. We show that the concept of reactive motion in this context is based on
a misunderstanding of the conservation of momentum in the presence of a gravitational field: such a reactive motion does not exist.
We show that the "mixed" equation used in some models is not physically consistent either. We examine why consideration of total momentum,
that is air and condensate combined, has been misleading in the search for valid equations of motion in the presence of phase transitions.
\end{abstract}

\section{Introduction}

The equation of motion for moist air in the presence of phase changes remains controversial in both
meteorological and multiphase flow literature \citep{young95,drew98,oo01,ba02,brennen05}.
\citet{young95} for example reviewed this subject and highlighted a number of inconsistencies in treatment and implied physical relationships.
In the atmospheric sciences this problem received attention in the works of \citet{oo01} and \citet{ba02}. But rather than resolving
the issues these authors offered contrasting equations.

The equation of motion for atmospheric air, without particles, can be written as
\begin{equation} \label{m0}
\rho \frac{\pt{\bf u}}{\pt t} + \rho ({\bf u} \cdot \nabla){\bf u} = {\bf F} + {\bf F}_c,
\eeq
\beq\label{F}
{\bf F} \equiv -\nabla p + \nabla \cdot \sigma + \rho {\bf g},\,\,\,{\bf F}_c = 0.
\end{equation}
Here ${\bf u}$ is air velocity, $\rho \equiv \rho_d + \rho_v$, $\rho_d$ and $\rho_v$ are moist air, dry air and water vapor density (kg~m$^{-3}$), $p$ is air pressure,
$\sigma$ is the viscous stress tensor, $\bf g$ is acceleration of gravity.
\citet{ba02} concluded that condensation causes an additional term ${\bf F}_c \ne 0$ to appear on the right-hand side of Eq.~(\ref{m0}):
\begin{equation} \label{Fc}
{\bf F}_c = \rho_l {\bf g} + {\bf F}_r,\,\,\,{\bf F}_r \equiv ({\bf u}_l - {\bf u}) \dot{\rho}.
\end{equation}
Here $\rho_l \equiv N_l \overline{m}$ is the density of condensate particles in the air, $N_l$ is the concentration of condensate particles
in the air, $\overline{m}$ is their mean mass, ${\bf u}_l$ is their velocity and $\dot{\rho} < 0$ (kg~m$^{-3}$~s$^{-1}$)
is condensation rate (the rate at which the moist air mass is diminished by phase transitions).
In the formulation of \citet{oo01} ${\bf F}_r = 0$.

\citet{ba02} mentioned the disagreement with \citet{oo01} but did not identify either its cause or implications.
The controversy persisted. Recently \citet{co11} reviewed the fundamental equations describing moist atmospheric dynamics and noted that
the correct way to consider momentum during phase changes remains poorly understood.
While some authors have argued that ${\bf F}_r$ in (\ref{Fc}) is usually small compared to the typical values of $\rho_l {\bf g}$ \citep{monteiro07},
they failed to rule out that under some circumstances it may be large. Indeed \citet{co11} concluded that this term
was of potential importance and warranted further study. \citet{co11} offered several illustrations explaining how the phenomenon is
manifested: according to these authors as the droplets begin to fall and thus gain a downward velocity, the remaining air gains an upward velocity so
that the combined momentum of air plus droplets is conserved \citep[][Fig. 2.2]{co11}.

However, any explanation based on the notion that the combined momentum of air plus droplets is conserved runs counter
to established physical principles.
Consider a rocket expelling fuel: it will experience acceleration proportional to the fuel velocity relative to itself
and mass ejected per unit time, because the total momentum of the system (rocket plus fuel) remains conserved. The expelled fuel acquires a
non-zero velocity relative to the rocket because of the {\it internal} forces generated when the fuel burns within the rocket.
Now consider a droplet of moisture forming in moist air:  unlike the rocket fuel, condensate particles in the atmosphere acquire a non-zero
velocity relative to the vapor from which they form due to an {\it external} force, gravity, imposed on them by a third body: the Earth.
Such acceleration causes no compensatory motion in the air.
(As another example, consider a block of ice melting on a table made of open mesh.  The block does not accelerate upwards as
the melt water streams down.) The momentum of the system is again conserved, as physics dictates, but here ``the system" includes the air,
the droplets and the entire bulk of the planet Earth.  The drop falls towards the Earth, and the Earth falls towards the drop.

Why might a reactive motion term appear in the equation of motion of moist atmospheric air? Here we review its origins.
In Section 2 we review some basic concepts showing the physical meaning of the reactive motion term and the associated pitfalls in
its formulations. In Section 3 we show that many derivations (including those in the leading textbooks on multiphase flow) appear based
on an incorrect formulation of Newton's second law when applied to systems of variable mass.
We then show that the reactive motion term in (\ref{Fc})
appeared in the work of \citet{ba02} as the result of an apparent mathematical error. We show that the omission of the reactive motion term by \citet{oo01} was correct even
though the reasoning behind it was not. We then consider Newton's third law as it applies to the interaction
between the air and droplets. We show that the equation used in many nonhydrostatic models of moist atmospheric
circulation is physically incorrect.
In the concluding section we discuss the search for valid equations of motion in the presence of phase transitions and explain why consideration of
total momentum (the so-called momentum equation) is unhelpful.

\section{Dynamics of bodies with variable mass}

In this section we outline the basic physical relationships governing the motion of bodies of variable mass.  This understanding
provides a foundation for the following sections where we identify the physical relationships governing moist atmospheric dynamics.

\subsection{Newton's second law}
Newton's second law for a body of mass $m$ and velocity $\bf u$ has the form
\beq\label{N}
m\frac{d \bf u}{dt} ={\bf f},
\eeq
where ${\bf f}$ represents the sum of all forces acting on the body.
This equation is valid for the case when the body mass changes with time,
$m = m(t)$, such that for the rate of momentum change we have
\beq\label{mom}
\frac{d (m{\bf u})}{dt} ={\bf f} + {\bf u}\frac{dm}{dt}.
\eeq

For $dm/dt = 0$ Eqs.~(\ref{N}) and (\ref{mom}) coincide, such that Eq.~(\ref{mom}) is referred
to as Newton's second law in many textbooks. However, a common misunderstanding of Newton's second law consists in putting variable mass $m(t)$ under the sign of differentiation in (\ref{N}),
see, for example, the discussions provided by \citet{pla92} and \citet{irschik04}:
\beq\label{merr}
\frac{ d (m \bf u)}{dt} ={\bf f},
\eeq
\beq\label{Nerr}
m\frac{d \bf u}{dt} ={\bf f} - {\bf u}\frac{dm}{dt}.
\eeq
The resulting equation of motion (\ref{Nerr}) is flawed:
it violates Galilean invariance by producing different acceleration in different inertial frames of reference.
An illustration follows. Consider a block of ice of mass $m$ that is melting in a warm room, $dm/dt < 0$, in the absence of net forces, ${\bf f}=0$.
Since velocity ${\bf u} = 0$, we have from (\ref{Nerr}) $m d{\bf u}/dt =0$: no acceleration, the ice block remains motionless
on the table.
Now consider this block of ice travelling in a train that moves with constant velocity ${\bf u} \ne 0$. From (\ref{Nerr}) we  obtain
$m d{\bf u}/dt = -{\bf u} dm/dt \ne 0$. Apparently the ice block accelerates within the train due to its melting.  Clearly this result is absurd and
the equation is incorrect. As we discuss below (see Section 3b) incorrect equations analagous to (\ref{merr}) and (\ref{Nerr})
when applied to particular constituents of the air and condensate mixture can be found in the literature on multiphase flow.

\subsection{Reactive motion and momentum conservation}

Consider our rocket: it has mass $m$ and is moving at velocity ${\bf u}$ in a free space (no external forces). It now expels some fuel
with mass $-\Delta m$ ($\Delta m < 0$) and velocity ${\bf u}_f$. The rocket's mass becomes
$m + \Delta m$ while its velocity becomes ${\bf u} + \Delta {\bf u}$.
As the total momentum of the system is conserved, we have
\beq\label{mc}
m {\bf u} = (m + \Delta m) ({\bf u} + \Delta {\bf u}) - \Delta m {\bf u}_f,\,\,\,\,
(m + \Delta m) \Delta {\bf u} = \Delta m ({\bf u}_f - {\bf u}).
\eeq
Considering the limit $\Delta m \to 0$ provides a differential equation for the dependence of our
rocket's velocity on its mass:
\beq\label{rm}
m\frac{d{\bf u}}{dm} = ({\bf u}_f - {\bf u}).
\eeq
If mass $m$ is a smooth function of time such that $(d{\bf u}/dm) (dm/dt) = d{\bf u}/dt$ we find
from (\ref{rm}) how the rocket's velocity depends on time:
\beq\label{rmt}
m\frac{d{\bf u}}{dt} = \dot{m}({\bf u}_f - {\bf u}),\,\,\,\dot{m} \equiv \frac{dm}{dt},
\eeq
this is the equation for our rocket's reactive motion.
Comparing (\ref{rmt}) and (\ref{N}) we can see that the rocket experiences a force equal to ${\bf f} = \dot{m}({\bf u}_f - {\bf u})$.

Now let us formally apply the law of momentum conservation to a system of two bodies with constant total mass $m = m_1 + m_2$
and velocities ${\bf u}_1$ and ${\bf u}_2$:
\beq\label{mc2}
\frac{d (m_1 {\bf u}_1 + m_2 {\bf u}_2)}{dt} = 0,
\eeq
\beq\label{mass}
\frac{dm_1}{dt} = - \frac{dm_2}{dt} \equiv \dot m_1.
\eeq
\beq\label{mot1}
m_1\frac{d{\bf u}_1}{dt} + m_2 \frac{d{\bf u}_2}{dt}= \dot{m}_1({\bf u}_2 - {\bf u}_1).
\eeq
Considering jointly (\ref{mot1}) and (\ref{rmt}) where in the latter $m = m_1$, ${\bf u} = {\bf u}_1$ and ${\bf u}_f = {\bf u}_2$, we find
that
\beq\label{zero}
\frac{d{\bf u}_2}{dt} = 0
\eeq
for $m_2 \ne 0$ in (\ref{mot1}).
This result is physically meaningful: as soon as the fuel
has acquired velocity ${\bf u}_2 = {\bf u}_f$ it is expelled. Since then it is no longer acted upon by any forces and continues to travel with a constant
 velocity. However, this result cannot be derived from (\ref{mc2}) without considering the equation of motion (\ref{rmt}).
Though the latter equation was also
obtained from the conservation of momentum, it contains more information than Eq.~(\ref{mc2}).

Note that equation~(\ref{mc2}) implies a restrictive assumption. It assumes that the fuel exiting the rocket at any given moment has the same
velocity ${\bf u}_2$ as all the fuel previously expelled (which has a total mass $m_2$). This is only correct if ${\bf u}_2 = {\bf u}_f$ is constant and
external forces are absent, such that the expel/exhaust maintains a constant velocity upon its exit from the rocket. But in the more general case where an
external force such as gravity is acting on the system, the expelled fuel can change velocity with time. Its mean velocity will then be unrelated
to that at the moment of expulsion. In contrast to Eq.~(\ref{mc2}), equation (\ref{rmt}) relates acceleration of the rocket to the velocity of the fuel
at expulsion irrespective of what happens to it afterwards. In this equation ${\bf u}_f$ is a free parameter.

Consider another example in which mass is gained. A train carriage is moving freely with velocity ${\bf u}$ along a long motionless platform,
from which people are throwing stones onto the carriage. We neglect the velocity components that are perpendicular to the carriage's
velocity, so for all bodies on the platform including the platform itself we have ${\bf u}_f = 0$.
According to (\ref{rmt}) the carriage will undergo a negative acceleration $du/dt = -(\dot m/m) u < 0$ because of its growing
mass $\dot{m} > 0$. If however people are running along the platform such
that the stones that they throw onto the carriage have the same velocity as the carriage, the carriage will not accelerate
despite the mean velocity of all bodies on the platform (the platform, the people and the stones) will still be different
from the velocity of the carriage.
These simple examples illustrate that the dynamics of reactive motion are determined by local velocities where the transition of mass from one body
to another takes place (i.e. the rocket's exhaust or the border between the carriage and the platform).
As we discuss below (see Section 3c) the neglect of this principle led to the incorrect derivation of \citet{ba02}.

\subsection{Newton's third law}

Equation (\ref{mc2}) has another drawback. It considers the system as composed of only two physical objects: one of mass $m_1$
moving with velocity ${\bf u}_1$ and another of mass $m_2$ moving with velocity ${\bf u}_2$. If this were true, Eq.~(\ref{mot1}) would
violate Newton's third law. Indeed, if there are no external forces acting on a system that consists of two bodies,
the only forces acting between them must be equal in magnitude and opposite in direction.
Then, according to Eq.~(\ref{N}), for the equations of motion of such bodies we must have
\beq\label{N3}
m_1 \frac{d{\bf u}_1}{dt} = {\bf f}_{12},\,\,m_2 \frac{d{\bf u}_2}{dt} = {\bf f}_{21},\,\,\,
{\bf f}_{12} + {\bf f}_{21} = 0,
\eeq
where the last equation expresses Newton's third law.
However, from (\ref{rmt}) and (\ref{zero}) we note that
${\bf f}_{12} = \dot{m}_1({\bf u}_2 - {\bf u}_1)$ and ${\bf f}_{21} = 0$, such that ${\bf f}_{21} \ne - {\bf f}_{12}$.
If, on the other hand, we consider (\ref{mot1}) independently of (\ref{rmt}) then from (\ref{N3}) and (\ref{mot1})
it follows that
\beq\label{dotm0}
{\bf f}_{12} + {\bf f}_{21} = \dot{m}_1 ({\bf u}_2 - {\bf u}_1) = 0.
\eeq
In other words, considering jointly Newton's third law as in (\ref{N3})
and Eq.~(\ref{mc2}) apparently yields an incorrect result: $\dot{m}_1 = 0$ at ${\bf u}_1 \ne {\bf u}_2$.

The controversy is resolved by noting that in reality there are {\it three}
objects that need to be distinguished, the third being the fuel
that is in the process of acceleration within the rocket. This fuel has velocity ${\bf u}_3$ intermediate between ${\bf u}_1$
and ${\bf u}_2$. The rocket does not impose any force on the fuel that has been already expelled (it has total mass $m_2$):
${\bf f}_{12} = {\bf f}_{21} = 0$.
The rocket only interacts with a force ${\bf f}_{13} = m_1 d{\bf u}_1/dt$ with the fuel that is accelerating.
By assuming that mass $\Delta m$ of this fuel is infinitely small in (\ref{rm}) and (\ref{rmt}), $\Delta m \to 0$ when $\Delta t \to 0$,
where $\Delta t$ is the period of time during which fuel mass $\Delta m$ was lost, we have required
that this fuel undergoes a finite change of velocity -- from ${\bf u}$ to ${\bf u}_f$ --
over an infinitely small period of time $\Delta t \to 0$, i.e. that it has an infinite acceleration.
So the accelerating fuel has a constant zero mass $m_3 \to 0$ and infinite acceleration $d{\bf u}_3/dt \to \infty$, such that their product $m_3 d{\bf u}_3/dt = {\bf f}_{31} =
-\dot{m}_1({\bf u}_2 - {\bf u}_1)$ is finite.  Upon exit from the rocket with velocity ${\bf u}_2$ the fuel is reclassified as belonging
to the second body. The instantaneously expelled fuel does not interact with the fuel that has been expelled previously, ${\bf f}_{32} = {\bf f}_{23} = 0$.
In this case Eq. (\ref{mc2}) is equivalent to ${\bf f}_{13} = - {\bf f}_{31}$.
In other words, Eq. (\ref{mc2}) is
brought in agreement with Newton's law by recognizing that there are three essential parts of the rocket-and-fuel system.

In the general case when there are external forces acting on a constant mass system
we can write Newton's law (\ref{N}) for any $k$-th of the system's components
as
\beq\label{Nk}
m_k\frac{d {\bf u}_k}{dt} ={\bf f}_{k} + \sum_{{\genfrac{}{}{0pt}{}{i}{(i \ne k)}}} {\bf f}_{ki},
\eeq
and the momentum conservation equation, cf. (\ref{mc2}), for
the system as a whole as
\beq\label{mck}
\sum_k \frac{d(m_k {\bf u}_k)}{dt} = \sum_k {\bf f}_{k}.
\eeq
where ${\bf f}_{k}$ is an external force (e.g. gravity) acting on the $k$-th component
and ${\bf f}_{ki}$ is the internal force imposed on the $k$-th component by the $i$-th component of the system.
If the system is considered as composed of only two components ($k = 1,2$), such that Newton's third law
reads ${\bf f}_{12} + {\bf f}_{21} = 0$, then Eqs.~(\ref{Nk}) and (\ref{mck}) yield the same incorrect
result $\dot{m}_1 ({\bf u}_2 - {\bf u}_1) = 0$, see (\ref{dotm0}). As we discuss below (see Section 3c)
it was this incorrect premise, ${\bf f}_{12} + {\bf f}_{21} = 0$, applied to the system of air and condensate
that led \citet{oo01} to conclude that the reactive motion term in (\ref{m0}) is absent. While as we discuss below this result is correct,
\citet{oo01} obtained it through flawed reasoning.

\section{Motion of moist air}

\subsection{In the absence of phase transitions}
Newton's second law for a moving fluid that occupies a material control volume $\tau$
(within which mass is conserved) can be written as follows \citep[][Eq.~3.1.11]{ba67}:
\beq
\label{N1}
{\bf f}= \frac{d}{dt}\int_\tau \rho {\bf u} d\tau = \int_\tau {\bf A} d\tau,\,\,\,
{\bf A} \equiv \rho \frac{D\bf u}{Dt} + {\bf u} \frac{D\rho}{Dt} + \rho {\bf u} (\nabla \cdot {\bf u}),
\eeq
\beq\label{Du}
\frac{D }{Dt} \equiv \frac{\partial}{\partial t} + ({\bf u} \cdot \nabla),
\eeq
where $\bf f$ is the sum of all forces acting on the fluid in the volume $\tau$.
In this Lagrangian representation the third term in $\bf A$ reflects the change of momentum density $\rho {\bf u}$
due to the change of the moving volume $\tau$:
because of $\nabla \cdot {\bf u} \ne 0$ some parts of the volume move with a larger velocity than the others, such that the
shape of the volume may change (see \citet[][]{ba67} for a discussion). Assuming that (\ref{N1}) is valid for any $\tau$, a differential
equation ${\bf A} = {\bf F}$ is obtained from (\ref{N1}), where ${\bf F}$ (\ref{F}) is the sum of forces acting on the fluid taken per unit volume:
the pressure gradient, viscous forces and gravity, $\int_\tau {\bf F} d\tau \equiv \bf f$.

Using the continuity equation
\beq
\label{cont0}
\frac{d}{dt}(\rho d\tau) = 0\,\,\,{\rm or}\,\,\,\frac{D\rho}{Dt} + \rho (\nabla \cdot {\bf u}) = 0,
\eeq
the equation of motion for the fluid (\ref{m0}) is obtained from (\ref{N1}):
\beq\label{mot0}
\rho \frac{D{\bf u}}{Dt} = {\bf F}.
\eeq

The Eulerian form of Eq.~(\ref{N1}) can be obtained by integrating ${\bf A}$ (\ref{N1}) over a spatially fixed (non-material) volume $V$ and using the Ostrogradsky-Gauss
theorem\footnote{Note that ${\bf A} = \rho \frac{\pt \bf u}{\pt t} + \rho ({\bf u} \cdot \nabla){\bf u} + {\bf u} \frac{\pt \rho}{\pt t}
+{\bf u}({\bf u} \cdot \nabla \rho) + \rho {\bf u} (\nabla \cdot {\bf u}) = \rho \frac{\pt \bf u}{\pt t} + {\bf u} \frac{\pt \rho}{\pt t} +
(\nabla \cdot {\bf u}) \rho {\bf u}$. On the other hand, from the Ostrogradsky-Gauss theorem we have in (\ref{eu})
$\oint_S \rho {\bf u} ({\bf u} d{\bf s}) = \int_V (\nabla \cdot {\bf u}) \rho {\bf u} dV$.}:
\beq\label{eu}
\int_V {\bf F} dV = \int_V {\bf A} dV = \frac{\pt}{\pt t} \int_V \rho {\bf u} dV + \oint_S \rho {\bf u} ({\bf u} d{\bf s}).
\eeq
Here the last term reflects the efflux of momentum from the considered volume enclosed by surface $S$ \citep[][p.~157]{young95,cr98}.

\subsection{In the presence of phase transitions}

A common derivation of the equations of motion in the presence of condensate starts by applying
Eq.~(\ref{N1}) in either integral or differential form
separately to each of the two components, see, e.g., \citet[Eq.~6.15]{drew98} and
\citet[Eq.~1.38]{brennen05}\footnote{This procedure was also offered by a reviewer of a previous draft of this work,
see details \href{http://www.bioticregulation.ru/common/pdf/ban/ban-history.pdf}{here}.}.

Since in a multiphase mixture with gas and condensate particles experiencing phase transitions
the mass of neither constituent is conserved -- though of course their total mass is -- two continuity equations are written in the differential form as follows,
see, e.g., \citet[][Eqs.~1.21~and~1.22]{brennen05} and \citet[][Eqs.~6.11~and~6.12]{drew98}:
\beq\label{contr}
\frac{D\rho}{Dt} + \rho (\nabla \cdot {\bf u}) = \dot{\rho},
\eeq
\beq\label{contl}
\frac{D\rho_l}{Dt_l} + \rho_l (\nabla \cdot {\bf u}_l) = -\dot{\rho},
\eeq
\beq\label{Dl}
\frac{D}{Dt_l} \equiv \frac{\partial}{\partial t} + ({\bf u}_l \cdot \nabla).
\eeq

Using these continuity equations the following equations of motion are obtained from Eq.~(\ref{N1}):
\beq
\label{air}
\rho \frac{D{\bf u}}{Dt} = {\bf F} + {\bf F}_{al} - \dot{\rho} {\bf u},
\eeq
\beq
\label{dro}
\rho_l \frac{D{\bf u}_l}{Dt_l}  = \rho_l {\bf g} + {\bf F}_{la} + \dot{\rho} {\bf u}_l.
\eeq

Here ${\bf F}_{al}$ and ${\bf F}_{la}$ are the additional friction forces exerted by the droplets on the air and vice versa,
respectively. These forces arise in the absence of phase transitions as well provided there are macroscopic particles  in the air.
These forces are formulated in different ways by various authors,
but their common property is that they sum up to zero:
\beq\label{3rd}
{\bf F}_{al} + {\bf F}_{la} = 0,
\eeq
see, for example, Eqs. 1.45 and 1.43 of \citet{brennen05} and Eqs. 7.13 and 6.24 of \citet{drew98}. By summing
(\ref{air}) and (\ref{dro}) using (\ref{3rd}) and assuming, as did \citet[][Eq.~5.13]{ba02}, that $D{\bf u}_l/Dt_l = 0$ in (\ref{dro}),
we obtain the equation of motion for moist air containing the reactive motion term ${\bf F}_r$ as in (\ref{Fc}).

However, as we discussed in Section 2a, applying Newton's law in the incorrect form (\ref{Nerr}) to bodies of variable mass
produces equations of motions that violate Galilelian invariance. This error is present in (\ref{air}) and (\ref{dro}):
similarly to (\ref{Nerr}), these equations contain the terms ${\bf u}\dot{\rho}$ and ${\bf u}_l\dot{\rho}$ that are proportional to the velocity
of the considered constituents. For example, consider
moist motionless air which initially does not contain condensate particles (such that ${\bf F}_{al} = 0$)
and is in hydrostatic equilibrium (such that $-\nabla p + \rho {\bf g} = 0$). At the moment when the vapor starts undergoing condensation,
$\dot{\rho} < 0$, the air will not accelerate. However, if we consider the same air in another inertial system
where ${\bf u} \ne 0$, then it will experience acceleration $\dot{\rho} {\bf u}/\rho \ne 0$.
Equations (\ref{air}) and (\ref{dro}) are thus as flawed as Eq.~(\ref{Nerr}).

A different physical approach is to write the budget of momentum for the mixture (air + droplets) as a whole
using the integral Eulerian form (\ref{eu}). In this approach total momentum of all the mass enclosed in
volume $V$ can change because of the efflux of matter from the volume and because of the sum of external forces acting on all the matter
enclosed within the volume \citep[][Eq.~38]{young95}:
\begin{equation} \label{re}
{\bf f}_{tot} \equiv \int_V {\bf F}_{tot}dV = \int_V\left[\frac{\pt (\rho {\bf u} + \rho_l {\bf u}_l)}{\pt t} +
(\nabla \cdot {\bf u})\rho{\bf u} + (\nabla \cdot {\bf u}_l)\rho_l{\bf u}_l \right]dV,
\eeq
\beq
{\bf F}_{tot} = {\bf F} + \rho_l {\bf g} = -\nabla p + \nabla \cdot \sigma+\rho {\bf g} + \rho_l {\bf g}.
\eeq
This approach does not rest on the incorrect Eqs.~(\ref{air}) and (\ref{dro}).
Eq.~(\ref{re}) is analagous to Eq.~(\ref{mck}).

Taking into account the continuity equations
(\ref{contr}) and (\ref{contl}) the following differential equation is obtained from (\ref{re})
by assuming that it remains valid for any volume \citep[e.g.,][Eq.~39]{young95}:
\beq
\label{D}
\rho \frac{D {\bf u}}{Dt} + \rho_l \frac{D {\bf u}_l}{Dt_l} = {\bf F} + \rho_l {\bf g} + {\bf F}_r,\,\,\,
{\bf F}_r \equiv ({\bf u}_l - {\bf u})\dot{\rho}.
\end{equation}
\citet[][Eq.~5.7]{ba02} obtained this equation having started from a differential form of Eq.~(\ref{re}).
Interpreting $D{\bf u}_l/Dt_l$ as the droplet's acceleration and assuming that droplets do not accelerate
over most part of their lifetime \citep[see][]{ba02}, the resulting equation of motion for moist air
$\rho D{\bf u}/Dt = {\bf F} + \rho_l{\bf g} + {\bf F}_r$, see (\ref{m0})-(\ref{Fc}), was obtained by \citet[][Eq.~5.18]{ba02}.

However, the transition from the integral to differential form in (\ref{re}) can only
be made when the functions under the integral sign lack singularities within the considered volume
such that du Bois-Reymond lemma is fulfilled \citep[e.g., see][p.~16]{drew98}.
Function $\dot \rho({\bf r})$, which describes the local process of phase transitions, does not satisfy this condition.
Phase transitions are localized on condensation nuclei and condensate particles surfaces.
If an elementary volume $dV$ encloses a droplet surface, in the limit $dV \to 0$ we have $\dot\rho({\bf r}) \to \infty$, assuming that surfaces occupy an
infinitely small volume. On the other hand, if $dV$ does not include a droplet surface, we have $\dot\rho({\bf r}) =0$ for any such $dV$.

Thus the rate of phase transitions taking place in a macroscopic volume $V_1$ that embraces one droplet of radius $r_s$
with droplet mass changing at a rate $dm/dt \equiv \dot m$ (kg~s$^{-1}$) can be
described as a delta function in a spherical coordinate system where $r = 0$ at the droplet center:
\beq\label{delta}
\dot\rho(r) = \sigma \delta(r - r_s),
\eeq
where $\sigma \equiv \dot m/(4\pi r_s^2)$ (kg~m$^{-2}$~s$^{-1}$) is the condensation rate per unit area of droplet surface.

In all practical applications Eq.~(\ref{D}) is averaged
over a finite volume. Averaging ${\bf u}\dot\rho$ in (\ref{D}) over $V_1 = (4/3)\pi R^3$, $R > r_s$, with use of (\ref{delta}) gives
\beq
\label{av}
\frac{4\pi}{V_1}\int_{V_1} {\bf u}\sigma \delta(r -r_s)r^2dr = {\bf u}_s \dot{\rho},
\eeq
where ${\bf u}_{s}$ is the mean air velocity at the surface of the droplet and $\dot \rho = \dot{m}/V_1$ is the volume-averaged condensation rate.

Because of viscosity the air velocity at the droplet surface coincides with droplet velocity (neglecting any minor effects due to droplet rotation
or increase in radius which due to symmetry make no impact on momentum) \citep[][Eq.~B.31]{cr98}.
We thus have ${\bf u}_{s} = {\bf u}_l$ in (\ref{av}), which means that the reactive motion term in (\ref{D}) equals zero.  It doesn't exist.

\citet{cr98} tackled the problem differently: without introducing a delta function
for the rate of phase transitions, they put $\dot \rho = 0$ for the continuous phase (the air) in the differential continuity equation (\ref{contr}),
see \citet[][Eq.~B.28]{cr98} and cf. Eq.~(\ref{contr}). The non-zero mean rate of phase transitions appeared
during the volume averaging calculated as the efflux of the continuous phase into the droplets\footnote{The
resulting equation  B.71 of \citet{cr98} for the momentum of the continuous phase is physically analagous to Eq.~(\ref{mom}) taking into account
that at the point where the phase transitions take place ${\bf u} = {\bf u}_s = {\bf u}_l$. This causes the term $\dot m {\bf u} = \dot m {\bf u}_l$
to appear in the right-hand side of Eq.~B.71 of \citet{cr98}, which cancels with the same term in the left-hand side of B.71 stemming from the
continuity considerations for the air momentum, i.e. ${\bf F}_r = 0$ in B.71.}.

\subsection{Derivations of \citet{ba02} and \citet{oo01}}

While quoting \citet{cr98} and noting that the flow variables in his approach, including Eq.~(\ref{D}),
represent volume averages, \citet{ba02} did not actually follow their procedure. \citet{ba02} replaced the correct term $F_r =
({\bf u}_l - {\bf u}_s)\dot{\rho} = 0$ by
$F_r = ({\bf u}_l - {\bf u})\dot{\rho} \ne 0$. This neglects that condensation is a surface-localized process, which occurs where the velocities of the
two phases coincide. This led to the appearance of the spurious reactive motion term ${\bf F}_r \ne 0$ in (\ref{D}).

Returning to our example with a moving train carriage and a motionless platform (Section 3b), phase transitions at the droplet surface
where ${\bf u} = {\bf u}_l$ can be compared
to the case when people are running along the platform with a velocity equal to that of the carriage, such
that the stones that they throw onto the carriage have the same velocity as the carriage. In this case the "phase transition"
(stone thrown onto the carriage) will not lead to any changes in momentum. It will only lead to a reclassification of the stone's mass
as belonging first to the platform, then to the carriage.

\citet{oo01} reached his result ${\bf F}_r = 0$ in (\ref{Fc}) differently. He wrote the equations of (vertical) momentum
separately for the air and the droplets in a form analagous to Eq.~(\ref{mom}), see \citet[][Eqs. 3.7 and 3.8]{oo01},
correctly adding terms $\dot \rho {\bf u}$ and $\dot \rho {\bf u}_l$ in the right-hand side of the respective equations.
These terms do not reflect any dynamics, but specify a reclassification
of mass to either part of the system. Eq.~3.7 of \citet{oo01}
for the vertical momentum of air is analagous to Eq.~6.30 of \citet{cr98} for the one-dimensional flow (but
note that in the latter equation it is taken into account that ${\bf u} = {\bf u}_l$ at the droplet surface where phase transitions occur.)

However, when specifying forces acting between the air and the droplets \citet{oo01}
made an erroneous assumption that we discussed in Section 2c, see Eqs.~(\ref{Nk}) and (\ref{mck}).
He assumed that these forces cancel. Thus his equations of motion lack any reactive motion by formulation,
see Eq.~(\ref{dotm0}). In reality, if the air had zero viscosity and a velocity different from that of the droplets where condensation occurred,
the interaction between the air and droplets could not be reduced to a pair of forces ${\bf F}_{la} = -{\bf F}_{al}$.
It would require inclusion of a third force describing the interactions
at the interface between the droplets and the air (analagous to fuel in the rocket). The final result of \citet{oo01} -- Eq.~(\ref{D}) with
${\bf F}_r = 0$ -- is correct, but apparently due to errors and good fortune.

\newpage
\subsection{Newton's third law and the equation of motion for the condensate}

Two further inconsistencies arise in the published derivations regarding the motion of moist air.
The first concerns the formulation of the so-called condensate loading term $\rho_l {\bf g}$ in (\ref{Fc}),
which is present in all models of moist atmospheric circulation that explicitly consider
the presence of condensate \citep[e.g.,][]{og63,da64,or82,ro87,dudhia93,oo01,ba02,br02,br09,co11}.
All derivations of (\ref{D}) that we have discussed make use of the assumption (\ref{3rd}), which means that
in a given volume droplets act on the air with the same force as the air acts on the droplets.
This statement can be often found in the literature as a version of Newton's third law
as applied to the moist air and the droplets \citep[e.g.,][]{og63,or82,pa00}.
This assumption is explicitly used in the derivations of \citet{drew98} and \citet{brennen05}.
It is also implicitly used in the derivations that start from the total momentum equation Eq.~(\ref{re})
as in \citet{young95} and \citet{ba02}. Indeed, in Eq.~(\ref{re}) total force ${\bf F}_{tot}$ acting on the matter enclosed
within a fixed (non-material) control volume $V$ does not include the impact of droplets that are located outside the considered volume.

This postulate (\ref{3rd}) is used to specify the unknown value of the force ${\bf F}_{al}$ by which the droplets act on the air.
Force ${\bf F}_{la}$ by which the air acts on the droplets can be found from the equation of droplet motion.
This equation was written by \citet[][Eq.~5.8a]{ba02} in the Eulerian form as
follows (note how it differs from (\ref{dro})):
\beq\label{drom}
\rho_l \frac{D{\bf u}_l}{Dt_l}  = \rho_l {\bf g} + {\bf F}_{la}.
\eeq
(Here in the notations of \citet{ba02} ${\bf F}_{la} = -\rho_l {\bf v}_l/\tau_{vl}$ and we neglected the effect of large-scale
pressure gradient on the droplet).
Assuming that the term in the right-hand side of Eq.~(\ref{drom}) is zero, we obtain the condensate loading term from (\ref{3rd}) and (\ref{drom}):
${\bf F}_{al} = -{\bf F}_{la} = \rho_l {\bf g}$.

In such an application of Newton's third law the "body" with which any given droplet interacts is defined as the air enclosed in the volume under consideration. However, an arbitrary chosen volume of air is in fact not a valid "body". For any droplet the true "body" with which it interacts is the continuous medium of the entire atmosphere
that is in hydrostatic equilibrium in the gravitational field of the Earth. Consider a droplet of linear size $r$ and mass $m$ falling at its terminal
velocity through an atmosphere without other droplets. The droplet's weight is compensated by an increase in hydrostatic surface pressure over a certain
region of linear size $L \gg r$ and area $L^2$  on the Earth's surface that is much larger than the area $r^2$ of the droplet's projection. Then the mean volume-specific
force ${\bf F}_{al}$ calculated in any intermediate unit volume $l^3$ surrounding the droplet, $r^3 \ll l^3 \ll L^3$, will be much smaller
than $\rho_l {\bf g} = m{\bf g}/l^3 \gg {\bf F}_{al}$. In this case the droplet's weight is negligible and can be neglected in the equation of motion of moist air averaged over this volume, ${\bf F}_c \approx 0$ in (\ref{Fc}).

Consider now a more realistic case with a large number of droplets falling simultaneously in the atmosphere. If their distribution is spatially non-uniform,
such that local density $\rho_l({\bf x})$ calculated for a unit volume $l^3$ changes significantly on a spatial scale $L \gg l$, for the volume-specific
force ${\bf F}_{al}$ imposed by droplets on the air enclosed in volume $l^3$ we will have ${\bf F}_{al} < \rho_l {\bf g}$ if $\rho_l > \overline{\rho}_l$
and ${\bf F}_{al} > \rho_l {\bf g}$ if $\rho_l < \overline{\rho}_l$, where $\overline{\rho}_l$ is the mean density of droplets in volume $L^3$.
An equality ${\bf F}_{al} \approx - {\bf F}_{la} = \rho_l{\bf g}$ could be fulfilled if the condensate was uniformly distributed in the atmosphere
such that its density $\rho_l$ would not vary in either horizontal or vertical direction. In the real atmosphere this is not the case -- e.g. consider
that most condensate is located where most condensation occurs: that is in the lower atmosphere \citep{pa12,jas13}. Thus the ubiquitous presence of the condensate
loading term in the scale-independent differential equation of moist air motion, see (\ref{m0})-(\ref{Fc}) and (\ref{Fco}), (\ref{Fc3}) below,
lacks justification.

The second inconsistency concerns the equation of motion for condensate particles.
Equation (\ref{drom}) containing the non-linear $({\bf u}_l \cdot \nabla){\bf u}_l$ term, see (\ref{Dl}), is identical in form to the Euler-Navier-Stokes
non-linear equations for fluids.  \citet{oo01} used the same non-linear formulation as \citet{ba02} in (\ref{drom}) to describe the acceleration of droplets.
However, motion of any particular droplet is described by the linear equations of Newton's second law; it is
independent of the presence of other droplets. There cannot be any general non-linear dynamic equation for the mean velocity ${\bf u}_l$ of droplets in
a unit volume. Equation (\ref{drom}) is {\it only} valid if all the droplets in the considered volume share the same velocity (see Appendix).
Thus, using a single velocity ${\bf u}_l \ne {\bf u}$ to describe the motion of droplets and using (\ref{drom})
all the newly formed droplets for which by definition ${\bf u}_l = {\bf u}$
were excluded from consideration by both \citet{oo01} and \citet{ba02}. This means that in their formulations droplets cannot form.

Additionally, while \citet{ba02} postulated that $D{\bf u}_l/Dt_l = 0$,  \citet{oo01} made a different assumption,
$D{\bf v}_l/Dt_l = 0$, where ${\bf v}_l \equiv {\bf u}_l - {\bf u}$ is droplet velocity relative to the air.
Thus \citet{oo01} assumed that droplet velocity relative to the air does not change along the droplet path.
However, since the droplet velocity relative to the air is largely dictated by the droplet's size,
assuming that droplets do not change their velocity along their path implies
that the rate of phase transitions, evaporation or condensation, is zero, $\dot \rho = 0$.
We conclude that using the non-linear Eulerian equation (\ref{drom}) for droplet motion does permit a physically
consistent description of phase transitions.

We will now discuss another equation of moist air motion found in the literature and how it compares
with the formulations of \citet{oo01} and \citet{ba02}.
Since
\beq\label{Dvl}
\frac{D{\bf u}_l}{Dt_l} = \frac{D{\bf v}_l}{Dt_l} + \frac{D{\bf u}}{Dt} + ({\bf v}_l \cdot \nabla) {\bf u},\,\,\,{\bf v}_l \equiv {\bf u}_l - {\bf u},
\eeq
the assumption of \citet{oo01}, $D{\bf v}_l/Dt_l = 0$, when used
in Eq.~(\ref{D}) with ${\bf F}_r = 0$ produces additional terms in Eq.~(\ref{m0}). Specifically, ${\bf F}_c$ in (\ref{m0}) becomes
\beq\label{Fco}
{\bf F}_c = \rho_l {\bf g} -\rho_l \frac{D{\bf u}}{Dt} - \rho_l ({\bf v}_l \cdot \nabla) {\bf u}.
\eeq
In the derivation of \citet{ba02} the last two terms in (\ref{Fco}) are absent. Commenting on
the work of \citet{oo01}, \citet{ba02} characterized them as spurious.

While some nonhydrostatic models of moist atmospheric circulation build
on the formulation of \citet{oo01} \citep[e.g.,][]{satoh03,satoh08},
in many other models neither Eq.~(\ref{Fc}) of \citet{ba02} nor
Eq.~(\ref{Fco}) of \citet{oo01} is used \citep[e.g.,][]{da64,ro87,dudhia93,br02,br09,co11}. Instead, in such models ${\bf F}_c$ in (\ref{m0}) is
given by \beq\label{Fc3}
{\bf F}_c = \rho_l {\bf g} -\rho_l \frac{D{\bf u}}{Dt}.
\eeq
with no justification. This gives rise to additional physical inconsistencies.

For example, \citet{br02}
while stating that they borrow their model equations from
\citet{ba02}, in reality put the reactive motion term ${\bf F}_r$ in (\ref{Fc}) equal to zero
and used Eq.~(\ref{m0}) with ${\bf F}_c$ given by (\ref{Fc3})
\citep[see][Eq.~1]{br02}. This can be reconciled with Eq.~(\ref{Fco}) of \citet{oo01} by noting
that in the model of \citet{br02} the droplets never fall down but float along
with the air, i.e. ${\bf v}_l = 0$. \citet{br02} in contrast to \citet{ba02}
assumed that ${\bf u}_l = {\bf u}$ and $D{\bf u}_l/Dt_l = D{\bf u}/Dt$ in (\ref{D}).
However, under such an assumption and considering that the force imposed by the air on the droplets is proportional
to their relative velocity ${\bf v}_l$, from (\ref{drom}) we obtain that ${\bf F}_{la} = 0$ and
$D{\bf u}_l/Dt_l = D{\bf u}/Dt = {\bf g}$. In other words, the air experiences free fall towards the Earth --
an absurd result.

In another model of \citet{br09}, which builds on the model of \citet{br02}, the droplet relative velocity ${\bf v}_l$ is
kept as a constant non-zero parameter in agreement with the proposition of \citet{oo01}.
But again in the equation of motion Eq.~(\ref{Fc3}) is used, which is equivalent to Eq.~(\ref{Fco}) with $({\bf v}_l \cdot \nabla) {\bf u} = 0$ \citep[][Eqs.~1-3]{br09}.
Since generally $({\bf v}_l \cdot \nabla) {\bf u} \ne 0$, this assumption is false.
Moreover, considering that in Eq.~(\ref{Dvl}) the major terms are those pertaining to the vertical dimension
we can estimate the relative importance of the term $Dw/Dt$ retained by \citet{br09} and
the term $W \pt w/\pt z$ that was discarded (here $W \equiv w_l - w$ is
the difference between the vertical velocities $w_l$ and $w$ of droplets and the air, respectively).
In a stationary case at $W \gg w$ and $Dw/Dt \sim w \pt w/\pt z$ the second, discarded, term can be significantly larger than the one retained.
To summarize, the presence of the $\rho_l D{\bf u}/Dt$ term in the models of moist atmospheric circulation is unjustified.

\section{Discussion}

We conclude with a general discussion of the problem of finding an equation of motion
in the presence of phase transitions. Conservation laws in the Eulerian form of the type
\begin{equation} \label{chi}
\frac{\partial \chi}{\partial t} + \nabla \cdot (\chi {\bf u}_\chi) = \dot{\chi},
\end{equation}
represent a general conservation relationship for an arbitrary property $\chi$ that is transported with velocity ${\bf u}_{\chi}$.
Such an equation presumes the knowledge of the volume source $\dot{\chi}$ of the considered property $\chi$. Without knowing this source
the general conservation laws (\ref{chi}) are a "tautologism" \citep{tru60}.

For momentum, such a source function is generally unknown. As we discussed in Section 2a, Newton's second
law does not describe the change of momentum in a general case of a variable mass. It only describes
the change of momentum if the system's mass is constant. Therefore, in the theoretical literature on
fluid dynamics, the Eulerian form of momentum equation (\ref{eu}) is obtained from Newton's equation of motion
and the continuity equation which ensures that the fluid mass is conserved \citep{ba67,land}.
The momentum equation for a fluid of type (\ref{chi}) is {\it not} an independent equation from which equations of motions could be derived.
Indeed, there are no grounds to assume {\it a priori} that the local change of momentum is determined by external forces acting on the system and
the efflux of matter from the system as per Eq.~(\ref{eu}) if we know that Newton's law is inapplicable to systems of variable mass.
How can we know that the efflux of matter from our considered non-material volume does not create additional
forces that would be absent when the efflux is zero? It is only from Lagrangian considerations
of Newton's law as in (\ref{N1}) that Eq.~(\ref{eu}) is justified.

However, once the fluid mass is not conserved (as in the presence of phase transitions), this approach loses validity.
We cannot apply Newton's law to the gas and the droplets in the Lagrangian approach (\ref{N1}), because their control
volumes move with different velocities, such that there is not a common control volume with a constant mass inside.
In this sense Eq.~(\ref{re}) is a postulate: we have postulated that the source function $\dot\chi$ for momentum
remains equal to the external force ${\bf F}_{tot}$ as is in the case of the fluid mass being conserved. Based on this postulate,
the attempts are made to derive the particular equations of motion for the air and condensate. However,
we could similarly postulate a particular physically reasonable form for the equation of motion directly.
Indeed, as we discussed in the previous section, considerations of the total momentum equation do
not provide for a physically consistent picture of motion for both droplets and the air.

We have discussed the various approaches to define the equation of motion of moist air based
on several papers addressing the fundamental relationships of multiphase flow \citep{young95,drew98,oo01,ba02,brennen05,co11}.
We have shown that some of these treatments incorrectly apply
Newton's second and third laws as well as the volume averaging procedures.
One conclusion is that no reactive motion term is required, the momentum of air plus droplets in the terrestrial atmosphere is not conserved
in isolation from the Earth. The correct treatments have been published and cited \citep{cr98}, but this has not prevented errors from creeping into
the meteorological literature.  While the contrasting mathematical treatments of moist air have raised little comment until now we hope that our
clarifications have shed some light on their cause and more generally on the dynamics of moist atmosphere.

{\bf Acknowledgements.} We thank three anonymous commenters on an earlier draft of this manuscript. BLL thanks the University of California Agricultural Experiment Station for their
partial support.

{\clearpage}
\begin{appendix}
\section*{
\begin{center}
The Eulerian form of equation of droplet motion
\end{center}}
Consider a flux of macroscopic particles (raindrops, hailstones, etc.) that do not interact with each other
contained within a fixed volume. We can define the mean velocity ${\bf u}_l$ and mean acceleration ${\bf a}_l$ of particles
as
\begin{equation}
\label{a}
{\bf u}_l \equiv \frac{1}{n}\sum_{i=1}^{n} {\bf u}_{li},\,\,\,{\bf a}_l \equiv \frac{1}{n}\sum_{i=1}^{n} {\bf a}_{li},\,\,\,
{\bf a}_{li} \equiv \frac{d{\bf u}_{li}}{dt}.
\end{equation}
Here $n$ is the number of particles in an arbitrary small volume $V$ enclosing the considered point, ${\bf u}_{li}$ and ${\bf a}_{li}$
are velocity and acceleration of the $i$-th particle.

We will show that equation of motion for condensate particles
\begin{equation} \label{dr}
\rho_l \frac{D{\bf u}_l}{Dt_l} = \rho_l {\bf g} + {\bf F}_{la}.
\end{equation}
is generally incorrect.

This equation must conform with Newton's second law governing the motion of particles. For a droplet of mass $m_i$ it reads:
\begin{equation}
\label{Nl}
m_i\frac{d{\bf u}_{li}}{dt}= m_i {\bf g} + {\bf f}_{li}.
\end{equation}
Here ${\bf f}_{li}$ is the force exerted by the air on the $i$-th droplet.

Summing (\ref{Nl}) over $n$ considered particles and dividing the sum by volume $V$ we obtain:
\beq
\label{al}
\frac{1}{V}\sum_{i=1}^n m_i{\bf a}_{li} = \rho_l {\bf g} + {\bf F}_{la},\,\,\,\rho_l = \frac{1}{V}\sum_{i=1}^n m_i,\,\,\,
{\bf F}_{la} \equiv \frac{1}{V}\sum_{i=1}^n {\bf f}_{li}.
\eeq

Comparing Eqs.~(\ref{al}) and (\ref{dr}) we find that Eq.~(\ref{dr}) can be obtained from Eq.~(\ref{Nl}) only if
the following relationship holds:
\begin{equation}
\label{er0}
\frac{1}{V}\sum_{i=1}^n m_i{\bf a}_{li} = \rho_l\left[\frac{\partial {\bf u}_l}{\partial t} + ({\bf u}_l \cdot \nabla) {\bf u}_l \right] \equiv \rho_l \frac{D {\bf u}_l}{D t_l}.
\eeq

We will consider the simplest case when
all the particles have the same mass, $m_i = m$ and $\rho_l = nm/V$. Eq.~(\ref{er0}) then reduces to
\begin{equation}
\label{er}
{\bf a}_{l} = \frac{\partial {\bf u}_l}{\partial t} + ({\bf u}_l \cdot \nabla) {\bf u}_l \equiv \frac{D {\bf u}_l}{D t_l}.
\end{equation}

For the purposes of our demonstration we divide the particles in the flux
into two arbitrary types
with respect to dynamics: e.g., let us colour these types green and red.
As the particles do not interact, the flux of green particles
is not affected by the presence of red particles, and vice versa. Thus, if (\ref{er}) were generally true,
it must apply to the red and green fluxes separately as well as to the combined flux.

Mean velocity ${\bf u}_l$ and mean acceleration ${\bf a}_l$ of the flow as a whole (red+green) are
\begin{equation}
\label{ava}
{\bf u}_l = \alpha {\bf u}_{l1} + \beta {\bf u}_{l2},\,\,\,{\bf a}_l = \alpha {\bf a}_{l1} + \beta {\bf a}_{l2},\,\,\,\alpha + \beta = 1,
\end{equation}
where $\alpha$ and $\beta$ are the relative numbers of red and green particles in the total flow,
${\bf u}_{lk}$ and ${\bf a}_{lk}$ (\ref{a}) are the mean velocity and acceleration of green ($k=1$) and red ($k=2$) particles,
respectively.

Let relationship (\ref{er}) be satisfied separately for both types of particles, i.e. the following is true for $k=1,2$:
\begin{equation}
\label{er1}
{\bf a}_{lk} = \frac{\partial {\bf u}_{lk}}{\partial t} + ({\bf u}_{lk} \cdot \nabla) {\bf u}_{lk}.
\end{equation}

Putting (\ref{er1}) and (\ref{ava}) into (\ref{er}), we find that (\ref{er}) only holds for the combined flow if
\begin{equation}
\label{rav}
([{\bf u}_{l1} - {\bf u}_{l2}] \cdot \nabla )({\bf u}_{l1} - {\bf u}_{l2}) = 0.
\end{equation}
This is true if either ${\bf u}_{l1} = {\bf u}_{l2}$ or
$\nabla {\bf u}_{l1} = \nabla {\bf u}_{l2}$. These conditions apply only when all particles possess the same velocity.

Such conditions do not hold in a precipitating atmosphere where some
droplets are already falling at near terminal velocity
and others have only begun to form and have near zero motion with respect to the surrounding air. Droplets of the first type have approximately zero mean acceleration ${\bf a}_{l1} \approx 0 \ll {\bf g}$,
while droplets of the second type have acceleration ${\bf a}_{l2} = {\bf g}$.

We have shown that Eq.~(\ref{dr}) is {\it not} valid if
the condensate is represented by particles having different velocities in a given point.
For identical droplets that in any given point have equal velocities,
the non-linear Eq.~(\ref{dr}) can be obtained from the linear Newton's equation
$\rho_l d{\bf u}_l/dt = \rho_l {\bf g} + {\bf F}_{la}$ by changing
variables from time $t$ to distance ${\bf l}$: $d{\bf l} \equiv {\bf u}_l dt$,
$d{\bf u}_l/dt \equiv ({\bf u}_l \cdot d/d{\bf l}) {\bf u}_l \equiv ({\bf u}_l \cdot \nabla) {\bf u}_l$.
Partial derivative $\pt /\pt t$ does not arise in this derivation.

For example, consider condensate particles of equal mass that originate at one at the same height
$z = h$ and start falling at $t = 0$. Such condensate particles will have equal velocity in each point.
Location $Z$ of the droplet on its trajectory depends on time $t$ as
\begin{equation}
\label{z}
Z = h + \int_0^t u_l(t')dt',\,\,\,\frac{dZ}{dt} = u_l(t),\,\,\,dt = \frac{dZ}{u_l(Z)}.
\end{equation}
Changing variables from $t$ to $Z$ in (\ref{Nl}) with use of (\ref{z}) we can re-write Eq.~(\ref{Nl}) as
follows (index $i$ is omitted):
\begin{equation}
\label{du2}
a = u_l\frac{du_l}{dZ} = -g + \frac{f}{m},\,\,\,\frac{d}{dZ} \frac{u_l^2}{2} = - g +\frac{f}{m}.
\end{equation}
Eq.~(\ref{du2}) follows from the linear equation (\ref{Nl}) and represents the law of energy conservation.
Non-linearity of Eq.~(\ref{du2}) appeared when $t$ was replaced by $Z$ in (\ref{Nl}).
Coordinate $Z=Z(t)$ in (\ref{du2}) describes the position of droplet on its trajectory and is
unambiguously determined by the dynamics of the droplet (\ref{Nl}) and time $t$.

If force $f$ acting on a droplet at point $Z(t)=z$ is time-independent,
all droplets forming at $z = h$ will have one and the same velocity $u_l(z)$ at point $z$.
There will be a stationary spatial distribution of particle velocity $u_l(z)$ (even if
condensation rate and, hence, particle density vary in time). We can see that in this case Eq.~(\ref{er}) is fulfilled
taking into account that $\partial u_l/\partial t = 0$.
On the other hand, if force $f(z)$ depends explicitly on time and/or if height $h$ where the condensate
originates depends on time, then particles formed at different
times may arrive at a point $z$ with different velocities. In this case, as we have shown, Eq.~(\ref{er})
is invalid.

\end{appendix}

\bibliography{met-refs}
\end{document}